# Astronomical aberration in the theory of ether and the theory of relativity. A new original interpretation

Daniele Sasso

e_mail: **dgsasso@alice.it**

**Abstract**

A new interpretation of astronomical aberration is given in this paper and it is based on the new theory of reference frames. The new interpretation confirms the result of classical physics based on the concept of ether and doesn't confirm the result of Special Relativity based on the constancy of the speed of light. But in the new theory there is no need to suppose the existence of ether.

## 1. Introduction

Astronomical aberration is the angular difference with respect to the earth observer between the apparent direction, specified by telescope, of a supposed still celestial body, for example a star, and its true direction, because of motions of the earth. This study on astronomical aberration results from the consideration that the relations of astronomical aberration in classical physics and in Special Relativity are different. The object of this research is a critical examination of the physical phenomenon in order to demonstrate a final equation that represents a valid mathematical model of astronomical aberration. In our study we will compare the two relations and we will show our equation proves the cogency of the classical relation and the invalidity of the relativistic equation. This result doesn't have the object to prove the superiority of the classical physics on the Special Relativity but only that with respect to astronomical aberration the classical result is more reliable than the Special Relativity.

## 2. Astronomical aberration in classical Physics

In classical Physics when star is on the axis of the ecliptic (for $\Phi=90°$) the aberration constant is[1]

$$tg\Phi' = v/c \qquad (1)$$

where $v$ is the tangential velocity of the earth in its orbit around the sun, $c$ is the velocity of light with respect to ether and $\Phi'$ is the angle of aberration between the apparent direction and the real direction of star. In general when star isn't on the axis of the ecliptic (fig.1) the aberration constant is given[2] (for $\Phi>90°$ as in figure) by

$$tg\Phi' = \frac{v\sin\Phi}{c - v\cos\Phi} \qquad (2)$$

Equation (1) is derived from (2) for $\Phi=90°$. Whether equation (1) or equation (2) are obtained in classical physics supposing the existence of ether and the velocity of light $c$ is calculated with respect to the reference frame of ether (ether theory).

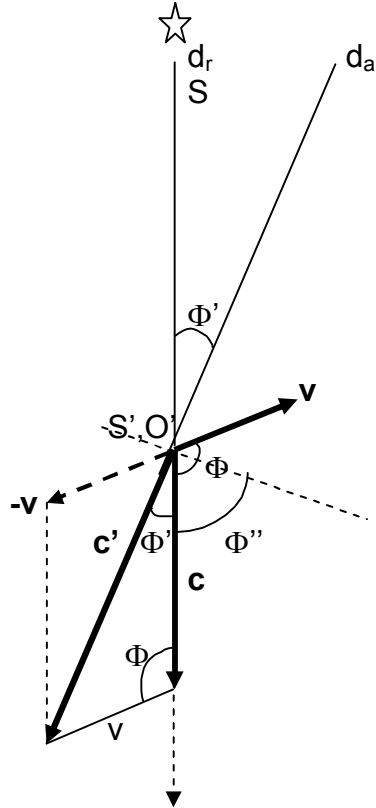

**FIG. 1.** The reference frame S is the physical system of the star supposed at rest; the reference frame S' is the physical system of the earth moving with velocity v in its orbital motion around the sun. The observer is in the origin O' of S'.
$\Phi$ is the angle between the real direction of the star and the velocity v of the earth with respect to the sun ; $\Phi''$ is the angle between the line star-observer and the perpendicular line to the apparent direction; $\Phi'$ is the angle of aberration.

### 3. Astronomical aberration in Special Relativity

In SR general aberration calculated by Einstein using the Lorentz space-time transformation [3] is

$$\cos\Phi' = \frac{\cos\Phi - \frac{v}{c}}{1 - \frac{v}{c}\cos\Phi} \qquad (3)$$

where $\Phi$ is the angle between the real direction of the star (line joining star-observer) and the velocity v of the earth observer with respect to the sun. When the star is on the axis of the ecliptic ($\Phi=90°$), as $\cos\Phi=0$, we have

$$\cos\Phi' = -v/c \qquad (4)$$

that is very different from the classical relation (1). A few scientists [1] have furnished a more credible interpretation of equation (3) assuming the expression

$$\cos\Phi'' = \frac{\cos\Phi - \frac{v}{c}}{1 - \frac{v}{c}\cos\Phi} \qquad (5)$$

where $\Phi''$ is the angle between the line star-observer and the perpendicular line to the apparent direction. In that case as $\Phi''=90-\Phi'$ (see fig. 1) we have in SR $\cos\Phi''= \sin\Phi'$ and therefore

$$\sin\Phi' = \frac{\cos\Phi - \frac{v}{c}}{1 - \frac{v}{c}\cos\Phi} \qquad (6)$$

and for $\Phi=90°$

$$\sen\Phi' = - v/c \qquad (7)$$

$$tg\Phi' = - \frac{1}{\sqrt{1 - v^2/c^2}} \frac{v}{c} \qquad (8)$$

Also in this case equation (8) derived from Einstein's equation (3) is different from equation (1) calculated in classical Physics. For very small angles (the measured angle of aberration is 20,47" seconds of degree) the two relations (1) and (8) give practically the same value of aberration, being v/c very small, but from the point of conceptual view they are very different. For this reason we want now to calculate the true expression of the aberration using the theory of reference frames[4].

### 4. The general law of astronomical aberration

#### 4.1 Application of the TR to astronomical aberration

The theory of reference frames (called briefly TR) is a new general theory whose the main conceptual novelty is methodological. The TR metodology is based on the Principle of Reference that for the analysis of any physical event sets out the existence of a preferred reference frame and observer. The preferred reference frame is the reference system where the physical event happens and the preferred observer is united with this reference system. The preferred reference system of TR has nothing in common with the absolute reference system of the classical physics that is one and coinciding with the ether; on the contrary the preferred reference system of TR changes with the physical system where the event happens. In order to study astronomical aberration the most important concept of TR is no doubt the assumption of two velocities of light: the physical velocity and the relativistic velocity. The physical velocity of light is constant and independent of the reference system and is the velocity measured in experiments; the relativistic velocity of light on the contrary changes with the reference system and is obtained by the theorem of vector composition of velocities.

## 4.2 Astronomical aberration

Astronomical aberration is caused by the apparent movement of stars that is the consequence of the movement with velocity v of the earth. If the star is on the axis of the ecliptic ($\Phi=90°$) this movement originates a small circle whose radius defines the aberration constant. If the star is at smaller latitude ($\Phi>90°$) it traces an elliptical small orbit whose larger semi-axis matches the radius of the previous circular orbit. By its nature the aberration is the same for all the stars.
Let us consider the figure 1 where S' is the earth reference system moving with velocity **v** with respect to the reference system S of the star supposed at rest. The observer is in the origin O' of the moving reference system S', **c** is the constant physical velocity of light with respect to the reference system S of the star and **c'** is the relativistic velocity of the same light with respect to the moving reference system S'. $d_a$ is the apparent direction of the star defined by measuring instrument, $d_r$ is the real direction of the star. $\Phi$ is the angle (declination of the star) between the real direction $d_r$ of the star and the earth velocity **v**, $\Phi'$ is the angle of aberration between the two directions $d_r$ and $d_a$.
Applying Carnot's theorem we have

$$v^2 = c^2 + c'^2 - 2cc'\cos\Phi' \qquad (9)$$
$$c'^2 = c^2 + v^2 - 2cv\cos\Phi \qquad (10)$$

Using equation (10) in equation (9) we obtain

$$v^2 = c^2 + v^2 + c^2 - 2cv\cos\Phi - 2cc'\cos\Phi'$$
$$c'\cos\Phi' = c - v\cos\Phi$$

and according to (10)

$$\cos\Phi' = \frac{1 - \dfrac{v\cos\Phi}{c}}{\sqrt{1 - 2\dfrac{v}{c}\cos\Phi + \dfrac{v^2}{c^2}}} \qquad (11)$$

If we set $\beta = v/c$ we have

$$\cos\Phi' = \frac{1 - \beta\cos\Phi}{\sqrt{1 - 2\beta\cos\Phi + \beta^2}} \qquad (12)$$

Equation (12) or the equivalent (11) is the general law of astronomical aberration. Let us see now some particular states:
  a. for $\Phi=0°$ and $\Phi=180°$ (longitudinal separation and approach of the earth observer with respect to the real direction of the star) we have $\cos\Phi'=1$ and $\Phi'=0°$ and therefore there isn't aberration;
  b. for $\Phi=90°$ (the star is on the axis of the ecliptic) we have $\cos\Phi=0$ and

$$\cos\Phi' = \frac{1}{\sqrt{1+\beta^2}} \qquad (13)$$

Equation (13) is the simplified expression of astronomical aberration.
We mark that our relation (13) appears different whether from the classical relation (1) or from the Einsteinian relation (8). By simple calculations we have from (13)

$$\tan \Phi' = \frac{\sin \Phi'}{\cos \Phi'} = \frac{\sqrt{1 - \cos^2 \Phi'}}{\cos \Phi'} = \beta = \frac{v}{c} \qquad (14)$$

Relation (13) calculated by us is therefore concordant with the classical relation (1) and in disagreement with the Einsteinian relation (8). Similarly it is easy to prove that equation (11) coincides with equation (2).

### 4.3 Annual and diurnal aberration

If in the relationships (12), (13) and (14) we assume v= "speed of the earth around the sun" we have annual aberration; if we assume v= "speed of the earth around its own axis" we have diurnal aberration.

## 5 Conclusion

Our research proves a new general equation regarding astronomical aberration. This equation is concordant with the classical relation and in disagreement with the Einsteinian relation. In fact Einstein's equation estimates the existence of effects of higher order

$$\tan \Phi' = \frac{1}{\sqrt{1 - v^2/c^2}} \frac{v}{c} \approx \frac{v}{c}\left(1 + \frac{1}{2}\frac{v^2}{c^2}\right) = \frac{v}{c}\left(1 + \frac{1}{2}\beta^2\right) \qquad (15)$$

These effects of higher order not have been never revealed in experimental work and moreover they have no physical explanation but follow only from the application of Lorentz's transformation that in TR is replaced by new space-time transformation.